\documentclass[twocolumn,showpacs,preprintnumbers,amsmath,amssymb]{revtex4}
\usepackage{graphicx}
\usepackage{dcolumn}
\usepackage{bm}

\newcommand\T{\rule{0pt}{2.6ex}}

\begin{document}
\title{Fragmentation processes in impact of spheres} \author{H.\ A.\
  Carmona$^{1,2}$, F. K. Wittel$^2$, F.\ Kun$^3$, and H.\ J.\
  Herrmann$^{2,4}$}

\affiliation{$^1$Centro de Ci\^encias e Tecnologia, Universidade
  Estadual do Cear\'a, 60740-903 Fortaleza, Cear\'a, Brazil}
\affiliation{$^2$ Computational Physics IfB, HIF, ETH,
  H\"onggerberg, 8093 Z\"urich, Switzerland}
\affiliation{$^3$Department of Theoretical Physics, University of
  Debrecen, P. O. Box:5, H-4010 Debrecen, Hungary}
\affiliation{$^4$Departamento de F\'isica, Universidade Federal do
  Cear\'a, 60451-970 Fortaleza, Cear\'a, Brazil}

\begin{abstract}
  We study the brittle fragmentation of spheres by using a
  three-dimensional Discrete Element Model. Large scale computer
  simulations are performed with a model that consists of agglomerates
  of many particles, interconnected by beam-truss elements. We focus
  on the detailed development of the fragmentation process and study
  several fragmentation mechanisms. The evolution of meridional cracks
  is studied in detail.  These cracks are found to initiate in the
  inside of the specimen with quasi-periodic angular distribution. The
  fragments that are formed when these cracks penetrate the specimen
  surface give a broad peak in the fragment mass distribution for
  large fragments that can be fitted by a two-parameter Weibull
  distribution. This mechanism can only be observed in 3D models or
  experiments. The results prove to be independent of the degree of
  disorder in the model.  Our results significantly improve the
  understanding of the fragmentation process for impact fracture since
  besides reproducing the experimental observations of fragment
  shapes, impact energy dependence and mass distribution, we also have
  full access to the failure conditions and evolution.
\end{abstract}
\pacs{46.50.+a, 62.20.Mk, 05.10.-a}
\maketitle

\section{Introduction}
Comminution is a very important step in many industrial applications,
for which one desires to reduce the energy necessary to achieve a given size
reduction and minimizing the amount of fine powder resulting from the
fragmentation process. Therefore a large amount of
research has already been carried out to predict the outcome of
fragmentation processes.  Today the mechanisms involved in the
initiation and propagation of single cracks are fairly well
understood, and statistical models have been applied to describe
macroscopic fragmentation \cite{hans:book, Astrom2006}. However, when
it comes to complex fragmentation processes with dynamic growth of
many competing cracks in three-dimensional space (3D), much less is
understood.  Today computers allow for 3D
simulations with many thousands of particles and interaction forces
that are more realistic than simple central potentials. These give a
good refined insight of what is really happening inside the system,
and how the predicted outcome of the fragmentation process depends on
the system properties.

Experimental and numerical studies of the fragmentation of single
brittle spheres have been largely applied to understand the elementary
processes that govern comminution \cite{GILVARRY1961a, GILVARRY1962,
  ARBITER1969, Andrews1998, Tomas1999, Majzoub2000, Chau2000,
  Salman2002, Wu2004, Schubert2005, Antonyuk2006, POTAPOV1995,
  Potapov1996, Potapov1997, Thornton1996, Kun1999, Thornton1999,
  Khanal2004, Behera2005, Herrmann2006}.  Experiments that were
carried out in the 60s analyzed the fragment mass and size
distributions \cite{GILVARRY1961a, GILVARRY1962,ARBITER1969} with the
striking result that the mass distribution in the range of small
fragments follows a power law with exponents that are universal with
respect to material, or the way energy is imparted to the system.
Later it became clear that the exponents depend on the dimensionality
of the object. These results were confirmed by numerical simulations
that were mainly based on Discrete Element Models (DEM) \cite{Kun1996,
  Kun1996a, Kun1999, Diehl2000, Astrom2000}. For large fragment
masses, deviation from the power law distribution could be modeled by
introducing an exponential cut-off, and by using a bi-linear or
Weibull distribution \cite{ODDERSHEDE1993, Potapov1996, Meibom1996,
  Lu2002, Cheong2004, Schubert2005, Antonyuk2006}. Another important
finding was, that fragmentation is only obtained above a certain
material dependent energy input \cite{ARBITER1969, Andrews1998,
  Andrews1999}. Numerical simulation could show that a phase
transition at a critical energy exists, with the fragmentation regime
above, and the fracture or damaged regime below the critical point
\cite{Kun1999,Thornton1999,Behera2005}.

The fragmentation process itself became experimentally accessible with the
availability of high speed cameras, giving a clear picture on the
formation of the fragments \cite{ARBITER1969, Andrews1998, Tomas1999,
  Majzoub2000, Chau2000, Salman2002, Wu2004, Schubert2005,
  Antonyuk2006}. Below the critical point, only slight damage can be
observed, but the specimen mainly keeps its integrity. Above but close to
the critical point, the specimen breaks into a small number of
fragments of the shape of wedges, formed by meridional
fracture planes, and additional cone-shaped fragments at the
specimen-target contact point. Way above the critical point,
additional oblique fracture planes develop, that further reduce the
size of the wedge shaped fragments.

Numerical simulations can recover some of these findings, but while
two-dimensional simulations cannot reproduce the meridional fracture
planes that are responsible for the large fragments \cite{POTAPOV1995,
  Potapov1997, Thornton1996, Kun1999, Khanal2004, Behera2005},
three-dimensional simulations have been restricted to relatively small
systems, and have not focused their attention on the mechanisms that
initiate and drive these meridional fracture planes
\cite{Potapov1996,Thornton1999}. Therefore, their formation and
propagation is still not clarified, although the resulting two to
four spherical wedged-shaped fragments are observed for a variety of
materials and impact conditions \cite{ARBITER1969, Majzoub2000,
  Khanal2004, Wu2004}. \citeauthor{ARBITER1969} \cite{ARBITER1969}
argued, based on the analysis of high speed photographs, that fracture
starts from the periphery of the contact disc between the specimen and
the plane, due to the circumferential tension induced by a highly
compressed cone driven into the specimen.  However, their experiments
did not allow access to the damage developed inside the
specimen during impact. Using transparent acrylic resin,
\citeauthor{Majzoub2000} \cite{Majzoub2000} observed damage initiation
at the border of the contact disc, but in their experiments plastic
flow and material imperfections may have a dominant role.

In this paper we present three-dimensional simulations of brittle
solid spheres under impact with a hard plate. With our simulations,
the time evolution of the fragmentation process and stress fields
involved are directly accessible. We have focused our attention on the
processes involved in the initiation and development of fracture, and
how they lead to different regimes in the resulting fragment mass
distributions. Our results can reproduce experimental observations on
fragment shapes, impact energy dependence, and mass distributions,
significantly improving our understanding of the fragmentation process
in impact fracture.
\section{Model and Simulation}
Discrete Element Models (DEM) have been successfully used since they
were introduced by \citeauthor{CUNDALL1979} to study rock mechanics
\cite{CUNDALL1979}. Applications range from static, to impact and
explosive loading, using elementary particles of various shapes that
are connected by different types of massless cohesive elements
\cite{POTAPOV1995,Kun1996a,Thornton1996,Kun1999,Thornton1999,
  Mishra2001,Thornton2004,Thornton2004a,Potyondy2004,Khanal2004,DAddetta2006,
  Carmona2007}. In general, Newton's equation governs the
translational and rotational motion of the elements, that concentrate
the whole mass. Forces and torques arise from element interactions,
from the cohesive elements, volumetric forces, and of course from
interaction with boundaries like walls.

Throughout this work we use a three-dimensional (3D) implementation of
DEM where the solid is represented by an assembly of spheres of two
different sizes. They are connected via beam-truss elements that
deform by elongation, shear, bending, and torsion.  The total force and
moment acting on each element consists of the contact forces resulting
from sphere-sphere interactions, $\vec{F}^{c} = \vec{F}^{ov} +
\vec{F}^{diss}$, the stretching and bending forces $\vec{F}^b =
\vec{F}^{elo} + \vec{Q}$ and moments $\vec{M^b}$ transmitted by the
beams attached.

\begin{figure}[ht]
  \centering
  \includegraphics[width=8.cm, bb= 0 0 462 552]{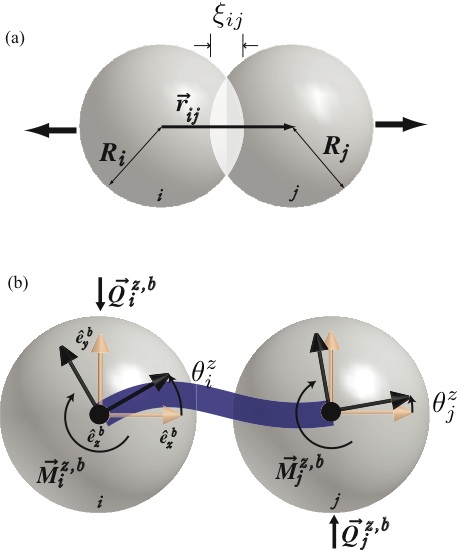}
  \caption{\label{fig:model}(a) Representation of the overlap
    interaction between two elements. (b) Typical deformation of a
    beam in the x-y plane, showing the resulting bending and shear
    forces and torques. The z-axis is perpendicular to the image.}
\end{figure}

The contact force has a repulsive term due to elastic interaction
between overlapping spherical elements, which is given by the Hertz
theory \cite{Landau1986} as a function of the material Young's modulus
$E^p$, the Poisson ratio $\nu^p$, and the deformation $\xi$. The force on
element $j$ at a distance $\vec{r}_{ij}$ relative to element $i$ (see
Fig.~\ref{fig:model}(a) ) is given by
\begin{equation} \label{eq:1} \vec{F}_j^{ov} =
  \frac{4}{3}\frac{E^p\sqrt{R^{eff}}}{(1-\nu^2)}\xi_{ij}^{3/2}\hat{r}_{ij},
\end{equation}
where the overlapping distance $\xi_{ij} = R_i + R_i -
\left|\vec{r}_{ij}\right|$ describes the deformation of the spheres,
$1/R^{eff} = 1/R_i + 1/R_j$, and
$\hat{r}_{ij}=\vec{r}_{ij}/\left|\vec{r}_{ij}\right|$.  The additional
terms of the contact force include damping and friction forces and
torques in the same way as described in Refs.
\cite{hans:book,HERRMANN1989,Kun1996a,Kun1999}.

The 3D representation of beams used in this work is an extension of
the two-dimensional case of Euler-Bernoulli beams described in Ref.
\cite{Poschel2005}.  In 3D, however, the total deformation of a beam
is calculated by the superposition of elongation, torsion, as well as
bending and shearing in two different planes. The restoring force
acting on element $j$ connected by a beam to element $i$ due to the
elongation of the beam is given by
\begin{equation} \label{eq:elo} \vec{F}_j^{elo} = -E^b A^b\varepsilon
  \hat{r}_{ij},
\end{equation}
where $E^b$ is the beam stiffness, $\varepsilon =
\left(\left|\vec{r}_{ij}\right| - l_0\right)/l_0 $, with the initial
length of the beam $l_0$ and its cross section $A^b$.

The flexural forces and moments transmitted by a beam are calculated
from the change in the orientations on each beam end,
relative to the body-fixed coordinate system of the beam
$(\hat{e}_x^b,\hat{e}_y^b,\hat{e}_z^b)$.  Figure~\ref{fig:model}(b)
shows a ty\-pi\-cal deformation due to rotation of both beam ends relative
to the $\hat{e}_z^b$ axis, with $\hat{e}_x^b$ oriented in the
direction of $\hat{r}_{ij}$.  Given the angular orientations
$\theta_i^z$ and $\theta_j^z$, the corresponding bending force
$\vec{Q}_j^{z,b}$ and moment $\vec{M}_j^{z,b}$ for the elastic
deformation of the beam are given by \cite{Poschel2005}:
\begin{subequations}\label{eq:bend}
  \begin{align}
    \vec{Q}_j^{z,b} &= 3 E^b I \frac{(\theta_i^z + \theta_j^z)}{L^2} \hat{e}_y^b,\label{benda}\\
    \vec{M}_j^{z,b} &= E^b I \frac{(\theta_i^z - \theta_j^z)}{L}
    \hat{e}_z^b + \left( \vec{Q}_i^{z,b} \times
      \left|\vec{r}_{ij}\right|\hat{e}_x^b \right),\label{bendb}
  \end{align}
\end{subequations}
where $I$ is the beam moment of inertia. Corresponding equations are written
for general rotations around $\hat{e}_y^b$, and the forces and moments
are added up. Additional torsion moments are added to consider a relative
rotation of the elements around $\hat{e}_x^b$:
\begin{equation} \label{eq:torsion} \vec{M}_j^{x,b} = -G^b I^{tor}
  \frac{(\theta_j^x - \theta_i^x)}{L} \hat{e}_x^b,
\end{equation}
with $G^b$ and $I^{tor}$ representing the shear modulus and
moment of inertia of the beams along the beam axis, respectively. The
bending forces and moments are transformed to the global coordinate
system before they are added to the contact, volume and walls forces.

Beams can break in order to explicitly model damage, fracture, and
failure of the solid.  The imposed breaking rule takes into account
breaking due to stretching and bending of a beam
\cite{HERRMANN1989,Kun1996,Kun1996a,DAddetta2001,Behera2005}, which
breaks if
\begin{equation}\label{pval}
  \left(\frac{\varepsilon}{\varepsilon_{th}}\right)^2 + \frac{\text{max}\left( |\theta_i|,|\theta_j| \right) }{\theta_{th}}\geq 1,
\end{equation}
where $\varepsilon = \Delta l/l_0$ is the longitudinal strain, and
$\theta_i$ and $\theta_j$ are the general rotation angles at the beam
ends between elements $i$ and $j$, respectively. Here $\cos \theta_i =
\hat{e}_x^{ib} \cdot \hat{e}_x^{b}$, where
$(\hat{e}_x^{ib},\hat{e}_y^{ib},\hat{e}_z^{ib})$ define the
$i-$particle's orientation in the beam body-fixed coordinate system,
similar calculation is performed to evaluate $\theta_j$.  Equation
(\ref{pval}) has the form of the von Mises yield criterion for metal
plasticity \cite{HERRMANN1989, Lilliu2003}.  The first part of Eq.~(\ref{pval})
refers to the breaking of the beam through stretching and the second
through bending, with $\varepsilon_{th}$ and $\theta_{th}$ being the
respective threshold values.  The introduced threshold values are
taken randomly for each beam, according to the Weibull distributions:
\begin{subequations}\label{eq:disorder}
  \begin{align}
    P(\varepsilon_{th}) &=
    \frac{k}{\varepsilon_o}\left(\frac{\varepsilon_{th}}{\varepsilon_o}\right)^{k-1}
    \exp\left[-\left(\frac{\varepsilon_{th}}{\varepsilon_o}\right)^{k}\right], \label{a}\\
    P(\theta_{th}) &=
    \frac{k}{\theta_o}\left(\frac{\theta_{th}}{\theta_o}\right)^{k-1}
    \exp\left[-\left(\frac{\theta_{th}}{\theta_o}\right)^{k}\right].\label{b}
  \end{align}
\end{subequations}
Here $k$ , $\varepsilon_o$ and $\theta_o$ are parameters of the model,
controlling the width of the distributions and the average values for
$\varepsilon_{th}$ and $\theta_{th}$ respectively. Low disorder is
obtained by using large $k$ values, large disorder by small $k$.
Disorder is also introduced in the model by the different
beam lengths in the discretization as described below.

The time evolution of the system is followed by numerically solving
the equations of motion for the translation and rotation of all
elements using a $6^{th}$-order Gear predictor-corrector algorithm, and the
dynamics of the rotations of the elements is described using
quaternions \cite{Rapaport2004,Poschel2005}.  The breaking rules are
evaluated at each time step. The beam breaking is irreversible, which
means that broken beams are excluded from the force calculations for
all consecutive time steps.
\subsection*{System formation and characterization}
Special attention needs to be given to the discretization in order to
prevent artifacts arising from the system topology, like anisotropic
properties, leading to non uniform propagation of elastic waves or
preferred crack paths.  In our procedure we first start with 27000
spherical elements that we initially place on a cubic lattice with
random velocities.  The element diameters are of two different sizes,
with $D_{2}=0.95 D_{1}$, that are randomly assigned, leading to more
or less equal fractions.  Once the elements are placed, the system is
left to evolve for 50000 time steps, using periodic boundary
conditions, in a volume that is about 8 times larger then the total
volume of the elements. This way we obtain truly random and uniformly
distributed positions.

To compact the elements, a centripetal constant acceleration field,
directed towards the center of the simulation box, is imposed. Due to
this field the elements form a nearly spherical agglomerate at the
center of the box. The system is allowed to evolve until all particle
velocities are reduced to nearly zero due to dissipative forces.

With the elements compacted, the next stage is to connect them by
beam-truss elements. This is achieved in our model through a Delaunay
triangulation of their positions.  As a consequence, not only
spherical elements that are initially in contact or nearly in contact
with each other are connected, but the resulting beam lattice is
equivalent to a discretization of the material using a dual Voronoi
tessellation of the material domain \cite{Lilliu2003, Bolander2005,
  Yip2006}.  After the bonds have been positioned, their Young's
moduli are slowly increased while the centripetal field is reduced to
zero. During this process the material expands to an equilibrium
state, reducing the contact forces. The bond lengths and orientations
are then reset so that no initial residual stresses are present in the
beam lattice. The final solid fraction obtained is approximately 0.65.
We have compared impact simulations of specimens compacted as
described above with specimens using random packings of spheres as
reported in Ref. \cite{Baram2005}, which have no preferential
direction in the packing process such as the one that could be imposed
by the centripetal field. No significant difference was found in the
simulation results, indicating that possible radially aligned
locked-in force chains are not relevant.

\begin{figure}[ht]
  \centering
  \includegraphics[width=8.5cm,bb= 0 0 488 488]{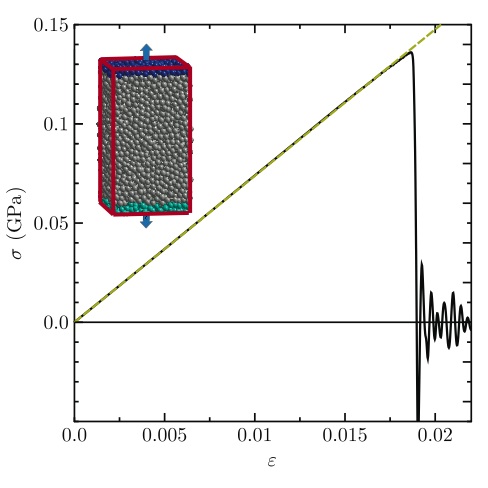}
  \caption{\label{fig:tens} Stress-strain curve for specimen under
    quasi-static loading. The inset shows the load geometry. Abrupt
    brittle fracture behavior can be observed at about $\varepsilon=0.019$.}
\end{figure}

Once the system is formed, the specimen is shaped to the desired
geometry by removing particles and beams that are situated outside the
chosen volume. The microscopic properties, namely the elastic
properties of the elements and bonds, as well as the bond breaking
thresholds, are chosen to attain the desired macroscopic Young's
modulus, Poisson's ratio, as well as the tensile and compressive
strength.  Table~1 summarizes the input values used in the simulations
presented in this paper. These were chosen to obtain macroscopic
properties close to the mechanical properties of polymers like PMMA,
PA, and nylon at low temperatures.  Figure~\ref{fig:tens} displays the
stress-strain curve measured by quasi-static, uni-axial tensile
loading of a bar, as depicted in the inset. The microscopic and
resulting macroscopic properties are resumed in Table~1, for a sample
size $(16 \times 8 \times 8 \text{ mm})$. The experiment is performed
by measuring the force required to slowly move the upper and lower
surfaces (see inset of Fig.~\ref{fig:tens}) at a constant strain rate
of $0.004 \text{ s}^{-1}$. The stress-strain curve is basically linear
until the strength is reached where rapid brittle fracture of the
material takes place. Oscillations in the broken specimen fractions
can be seen after the system is completely unloaded due to elastic
waves. The Young's modulus measured from the slope of the curve is
$7.4 \pm 0.5 \text{GPa}$, is presented along with other macroscopic
properties of the material in Table~1.

In order to simulate the impact of a sphere on a frictionless hard
plate, a spherical specimen with diameter $D=16 \text{ mm}$ is
constructed, and a fixed plane with Young's modulus $70 \text{ GPa}$ is
added to the simulation. The spherical specimen has a total of
approximately 22000 elements, with around 32 across the sample
diameter.  The contact interaction between the elements and the plate
is identical to the element-element contact interaction, only with
$\xi = R_i - r_{ip}$ , where $r_{ip}$ is the distance between the
particle center and the plate. The specimen is placed close to the
plate with an impact velocity $v_i$, in the negative z-direction,
assigned to all its composing elements. The computation continues
until no additional bonds are broken for at least $50 \mu s$.

\begin{figure}[ht]
  \centering
  \includegraphics[width=8.5cm, bb= 0 0 482 562]{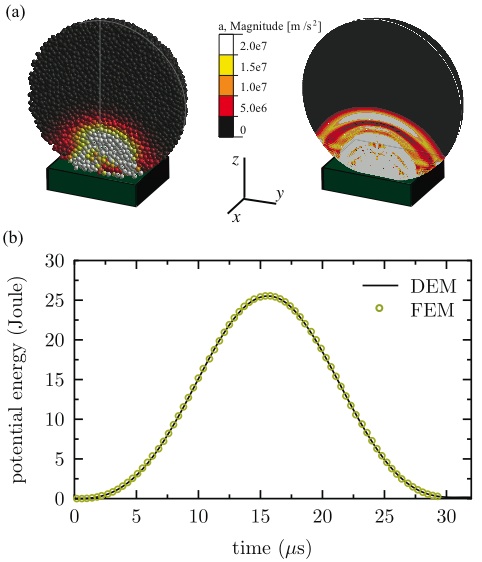}
  \caption{ \label{fig:demfem}(a) Comparison of deformations and
    shock-wave propagation obtained between DEM and FEM simulations
    for $v_i=117 \text{ m/s}$. (b) Time evolution of the elastic
    potential energy stored in the system for the same velocity
    obtained by DEM (solid line) and FEM (dashed line) simulations. }
\end{figure}

For comparative reasons we calculate the evolution of the stress field
using an explicit Finite Element (FE) analysis. The FE model is
composed of axisymmetric, linear 4-node elements with macroscopic
properties taken from the results of the DEM simulations (see
Table~1).  Along the central axis through the sphere and ground plate,
symmetry boundary conditions are imposed, the bottom of the target
plate is encastred and contact surfaces for the sphere and plate are
defined.  Figure~\ref{fig:demfem}(a) shows a comparison between the
impact simulation using our DEM model and a Finite Element Model
simulation.  In Fig.~\ref{fig:demfem}(a), the DEM elements are colored
according to the amplitude of their accelerations to show the
propagation of a longitudinal shock wave that was initiated at the
contact point. The wave speed can be estimated to be approximately
$2200 \pm 100 \text{m/s}$, which is consistent with the Young's
modulus of the material derived from Fig.~\ref{fig:tens} and its
density. The time evolution of the potential energy stored in the
system is compared in Fig.~\ref{fig:demfem}(b), showing excellent
quantitative agreement between the two models.

After the characterization of the system properties we allow for the
cohesive elements to fail in order to study the fragmentation
properties.

\begin{table}{l}\label{tab:properties}
\caption{Micro- and macroscopic material model properties.}

\begin{tabular}{l}
   Typical model properties (DEM):\\ 
   \hline
   \T
   Beams:\\
   \begin{tabular}{lllll}
     & stiffness~~~~~~~~~~~~~~~~~ & $E^b/G^b$~~~ & 6~~~~~~~~~~ & GPa\\
     & average length & $L$ & 0.5 & mm\\
     & diameter & $d$ & 0.5 & mm\\
     & strain threshold & $\varepsilon_{0}$ & 0.02 & -\\ 
     & bending threshold & $\theta_{0}$ & 3 & $^\circ$\\
     & shape parameter & $k$ & 3/10 & -\\
   \end{tabular} \\
   Particles:\\
   \begin{tabular}{lllll}
     & stiffness~~~~~~~~~~~~~~~~~ & $E^p$~~~~~~~ &3~~~~~~~~~~ & GPa\\
     & diameter& $D_{1}$ & 0.5 & mm\\
     & density & $\rho$ & 3000 & kg/m$^3$\\ 
   \end{tabular} \\
   Hard plate:\\
   \begin{tabular}{lllll}
     & stiffness~~~~~~~~~~~~~~~~~ & $E^w$~~~~~~~ &70~~~~~~~~~ & GPa\\
   \end{tabular} \\
   Interaction:\\
   \begin{tabular}{lllll}
     & friction coefficient~~~~ & $\mu$~~~~~~~~~ & 1~~~~~~~~~~ & -\\
     & damping coefficient & $\gamma_n$ & 0.25& s$^{-1}$\\ 
     & friction coefficient & $\gamma_t$ & 0.05 & s$^{-1}$\\ 
   \end{tabular} \\
   System:\\
   \begin{tabular}{lllll}
     & time increment~~~~~~~~ & $\Delta t$~~~~~~~ & 1e-8~~~~~~ & s\\
     & number of particles & $N^p$ & 22013 & -\\ 
     & number of beams & $N^b$ & 135948 & -\\ 
     & solid fraction &  & 0.65 & -\\ 
     & sphere diameter & $D$ & 16 & mm\\
   \end{tabular} \\ 
   \\
   Macroscopic properties (DEM):\\
   \hline
   \begin{tabular}{lllll}
     \T
     & system stiffness~~~~~~~~ & $E$~~~~~~~~ & $7.4\pm 0.5$ & GPa\\
     & Poisson's ratio & $\nu$ & 0.2 & -\\ 
     & density & $\rho$ & 1920 & kg/m$^3$\\ 
     & system strength & $\sigma_c$ & 110 & MPa\\ 
   \end{tabular} \\ 
   \\
   Comparison:\\
   \hline
   \begin{tabular}{lllll}
     \T
     & & DEM  & FEM &\\
     & longitudinal~~~~ & $2210\pm 100$~~  & $2270\pm 20$~~ & m/s\\
     & wave speed &  &  &\\\T
     & contact time & 31.4 &31.4& $\mu$s\\
   \end{tabular} \\
 \end{tabular}

\end{table}
 \section{Fragmentation mechanisms}
\begin{figure}[ht]
   \centering
   \includegraphics[width=8.5cm, bb= 0 0 470 208]{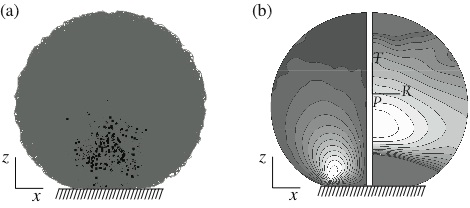}
   \caption{ \label{fig:diff_biaxial}Initial damage due to bi-axial
     stress state. (a) Vertical cut through the center of the sphere
     from the DEM simulation showing broken bonds represented by dark
     color. (b) Stress fields calculated with FEM model. Left side are
     shear stresses in global coordinates from 0 to -400MPa (black to
     white) while the right side shows circumferential stresses in
     local spherical coordinates with the center in the sphere center
     ranging from 0 to 130MPa (black to white).}
 \end{figure}

 In this section we explore the different fragmentation mechanisms in
 the order of occurrence and increasing energy input. The first yield
 that arises in the material is diffuse damage that occurs in the
 region above the contact disc. It can be seen from
 Fig.~\ref{fig:diff_biaxial}(a), that this damage region is centered
 in the load axis, at a distance approximately $D/4$ from the plane.
 
 We can see a strong correlation of the position of the diffuse
 cracking in the DEM results (Fig.~\ref{fig:diff_biaxial}(a)) with the
 location of a region with a bi-axial stress state in the x-y-plane
 and a superimposed compression in the z-direction, as calculated
 using FEM (Fig.~\ref{fig:diff_biaxial}(b)), also in agreement with
 experimental results reported in Ref.~\cite{Andrews1998} . This
 result, along with the one presented in Fig.~\ref{fig:demfem},
 suggests that the use of three-dimensional beams, as compared with
 the use of simple springs, despite of the reduced number of degrees
 of freedom in the breaking criterion, could recover quite well the
 influence of complex stress states in the crack formation in a more
 precise way.

\begin{figure}[ht]
   \centering
   \includegraphics[width=8.5cm, bb = 0 0 482 544]{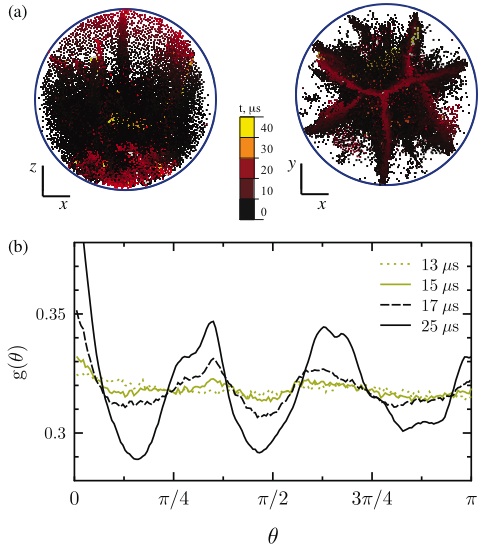}
   \caption{\label{fig:merid_cracks} (a) Colored dots display the
     positions of the broken bonds according to the time of breaking.
     (b) Angular distribution function of broken bonds as a function
     of the angular separation when their positions are projected into
     the plane perpendicular to the load axis.}
 \end{figure}

 As time evolves, meridional cracks start to appear.  The origination
 of this type of cracks is explored in Fig.~\ref{fig:merid_cracks},
 where we plot in Fig.~\ref{fig:merid_cracks}(a) the positions of the
 broken bonds in two different projections, showing well defined
 meridional crack planes that propagate towards the lateral and upper
 free surfaces of the specimen. In Fig.~\ref{fig:merid_cracks}(b) we
 plot the angular distribution of the broken bonds for different
 times.  Here $g(\theta)$ is the probability of finding two broken
 bonds with an angular separation $\theta$. Note that their positions
 are projected into the plane perpendicular to the load axis. The
 evident peaks in $g(\theta)$ are a clear indication that the cracks
 are meridional planes that include the load axis. In this particular
 case, the cracks are separated by an average angle of about 60
 degrees, and they become evident 13 to 15 $\mu$s after impact
 ($v_i=120 \text{ m/s}$).

 In order to understand what governs the orientation and angular
 separation of these meridional cracks we performed many different
 realizations with different seeds of the random number generator and
 impact points. For all cases the orientation of the cracks can
 change, but not their average angular separation.  We observe that
 for strong disorder (Eq.~(\ref{eq:disorder})), a larger amount of
 uncorrelated damage occurs, but the average angular separation of the
 primary cracks does not change. This suggests that the formation of
 these cracks arises due to a combination of the existence of local
 disorder and the stress field in the material, but does not depend on
 the degree of disorder.

 As we can see from the FE calculations and from the damage
 orientation correlation plot (Fig.~\ref{fig:merid_cracks}) inside of
 the uniform biaxial tensile zone, no preferred crack orientation is
 evidenced. Many microcracks weaken this material zone, decreasing the
 effective stiffness of the core. Around the weakened core the
 material is intact and under high circumferential stresses. It is in
 this ring shaped zone, that we observe to be the onset of the
 meridional cracks when we trace them back.  For increasing impact
 velocity we observe a decrease in the angular separation of crack
 planes and thus more wedge-shaped fragments. Therefore this fragment
 formation mechanism can not be explained by a quasi-static stress
 analysis.  The observation is in agreement with experimental findings
 and can be explained by the basic ideas of Mott's fragmentation
 theory for expanding rings \cite{MOTT46}.  Due to the stress release
 front for circumferential stresses, once a meridional crack forms,
 stress is released in its neighborhood; the fractured regions spread
 with a constant velocity and the probability for fracture in
 neighboring regions decreases.  On the other hand in the unstressed
 regions, the strain still increases, and so does the fracture
 probability along with it.  The average size of the wedge shaped
 fragments therefore is determined by the relationship between the
 velocity of the stress release wave and the rate at which cracks
 nucleate.  Thus the higher the strain rate, the higher the crack
 nucleation rate and the more fragments are formed. We measured the
 strain rate at different positions inside the bi-axially loaded zone,
 finding a clear correlation between impact velocity and strain
 rate. Even though we fragment a compact sphere and not a ring, when
 it comes to the formation of meridional cracks, we observe that they
 form in a ring shaped region and that Mott's theory can qualitatively
 explain the decrease of angular separation between wedge shaped
 fragments with increasing impact velocity.

 If enough energy is still available, some of the meridional plane cracks grow
 outwards and upwards, breaking the sample into wedge shaped
 fragments that resemble \emph{orange slices}. 

\begin{figure}[ht]
   \centering
   \includegraphics[width=8.5cm, bb = 0 0 482 226]{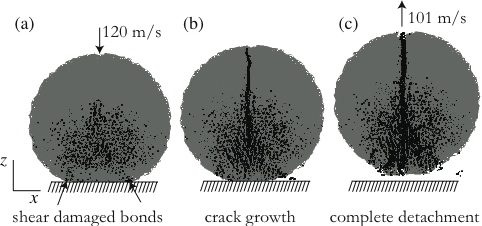}
   \caption{ \label{fig:cone} Vertical meridional cut of the sphere at
     different stages during impact, showing the separation of lower
     fragments. (a) Formation of a ring of broken bonds due to shear
     failure. (b) These broken bonds evolve into cracks that propagate
     inside the sample. (c) Finally these cracks lead to the
     detachment of the lower fragments.}
 \end{figure}

 As the sphere continues moving towards the plate, a ring of broken
 bonds forms at the border of the contact disc due to shear failure
 (compare Fig.~\ref{fig:cone}(a)). When the sphere begins to detach
 from the plate, the cone has been formed by high shear stresses in
 the contact zone (see Fig.~\ref{fig:diff_biaxial}(b) left) by a ring
 crack that was able to grow from the surface to the inside of the
 material under approximately 45$^\circ$ (Fig.~\ref{fig:cone}(b)). It
 detaches, leaving a small number of cone shaped fragments that have
 a smaller rebound velocity than the rest of the fragments due to
 dissipated elastic energy (Fig.~\ref{fig:cone}(c)).

\begin{figure}[ht]
   \centering
   \includegraphics[width=8.4cm, bb=0 0 480 232]{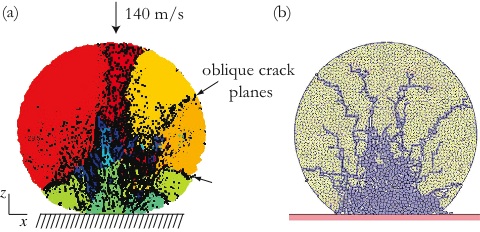}
   \caption{\label{fig:secondary} (a) DEM simulation at $v_i=140\text{
       m/s}$ exemplifying the secondary cracks. The bonds are colored
     according to the final fragment they belong to. (b) 2D
     simulations using polygons as elementary particles
     \cite{Behera2005}.}
 \end{figure}

 Oblique plane cracks may still break the large fragments further, if
 the initial energy given to the system is high enough. Therefore they
 are called \emph{secondary} cracks. Figure~\ref{fig:secondary}(a)
 shows a vertical meridional cut of a sample where these cracks can be
 seen. The intact bonds are colored according the final fragment they
 belong to.

 These secondary cracks are very similar to the oblique cracks
 observed in 2D simulations \cite{POTAPOV1995, Behera2005}. For
 comparison we show in Fig.\ref{fig:secondary}(b) the crack patterns
 obtained from a 2D DEM simulation that uses polygons as elementary
 particles.  In the 2D case, we observe a cone of numerous single
 element fragments and meridional cracks cannot be observed.

 \section{Fragmentation regimes}

\begin{figure}[ht]
   \centering
   \includegraphics[width=8.5cm, bb = 0 0  454 732]{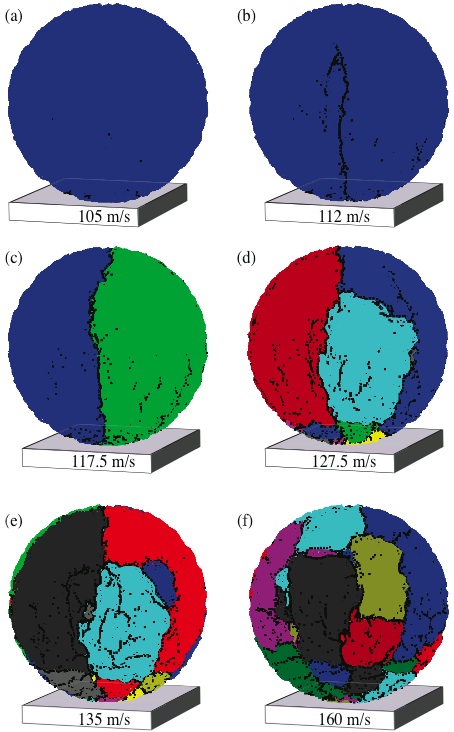}
   \caption{\label{fig:final_cracks} Front view of the reconstructed
     spheres, showing the final crack patterns at the surfaces for
     different initial velocities. Gray dots are placed at the positions
     of broken beams while different colors are chosen for different
     fragments.}
 \end{figure}

 The amount of energy necessary to fragment a material is a parameter
 that is very important for practical applications in comminution.  In
 fragmentation experiments two distinct regimes for damage and
 fragmentation can be identified depending on the impact energy: below
 a critical energy \cite{GILVARRY1962, Andrews1998, Thornton1999}
 damage takes place, while above fragments are formed.
 Figure~\ref{fig:final_cracks} compares the final crack patterns after
 impact with different initial velocities. The intact bounds are
 colored according to the final fragment they belong to, and gray dots
 display the positions of broken bonds. The fragments have been
 reassembled to their initial positions to provide a clearer picture
 of the resulting crack patterns.  For the smaller impact velocities
 it is possible to identify meridional cracks that reach the sample
 surface above the contact point, but fragmentation is not complete
 and one large piece remains (Figs.~\ref{fig:final_cracks}(a) and
 \ref{fig:final_cracks}(b)). We call these meridional cracks
 \textit{primary cracks}, since as one increases the initial energy
 given to the system, some of them are the first to reach the top free
 surface of the sphere, fragmenting the material into a few large
 pieces, typically two or three fragments with wedge shapes
 (Fig.~\ref{fig:final_cracks}(c)).  When we increase the initial
 energy secondary oblique plane cracks break the orange slice shaped
 fragments further (Fig.~\ref{fig:final_cracks}(d)). Additional
 increase in the impact velocity causes more secondary cracks and
 consequently the reduction of the fragment sizes
 (Figs.~\ref{fig:final_cracks}(e) and \ref{fig:final_cracks}(f)).

 The shape and number of large fragments resulting from the numerical
 model for smaller impact energies, as well as the location and
 orientation of oblique secondary cracks for larger energies, are in
 agreement with experimental findings \cite{Khanal2004, Schubert2005, Wu2004}.

 We can identify that for velocities smaller then a threshold
 value, the sample is damaged by the impact but not fragmented.  This
 threshold velocity for fragmentation has been found experimentally
 and numerically \cite{Andrews1998, Kun1999, Behera2005}.  In
 particular, it has been found from 2D simulations that a continuous
 phase transition from damaged to fragmented  outcome of
 impact fragmentation can be tuned by varying the initial energy
 imparted to the system \cite{Behera2005,Kun1999}.

\begin{figure}[ht]
   \includegraphics[width=8.5cm, bb = 0 0 488 488]{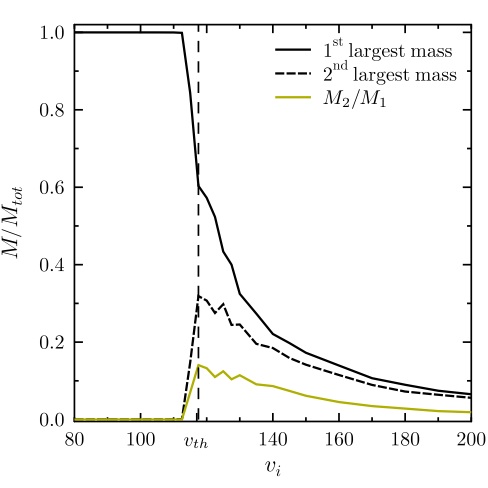}
   \caption{\label{fig:largest} The mass of first and second largest
     fragment and the average fragment mass as a function of the
     impact velocity.}
 \end{figure}

 Following the analysis in references \cite{Behera2005,Kun1999} the
 final state of the system after impact is analyzed by observing the
 evolution of the mass of the two largest fragments, as well as the
 average fragment size (shown in Fig.~\ref{fig:largest}). The
 average mass $M_2/M_1$, with $M_k = \sum_i^{N_f}{M_i^k - M_{max}^k}$
 excludes the largest fragment. It can be observed that below the threshold
 value $v_{th} = 115 \text{ m/s}$ the largest fragment has almost the
 total mass of the system, while the second largest is orders of
 magnitude smaller. This behaviour implies that for $v_i < v_{th}$ the
 system does not fragment, it only gets damaged.
 For velocities larger then $v_{th}$ the mass of the largest fragment
 decreases rapidly. The second largest and average fragment masses
 increase, having their maximum at $117.5 \text{ m/s}$ for this
 material strength.

 The results shown in Fig.~\ref{fig:largest} are in very good
 qualitative agreement with those obtained from simulations in
 different geometries and load conditions \cite{Kun1999, Behera2005,
   Wittel2005}, indicating that our model shows a phase transition
 from a damaged to a fragmented state.

 \section{Resulting fragment mass distribution}
\begin{figure}[ht]
   \includegraphics[width=8.5cm, bb = 0 0 488 800]{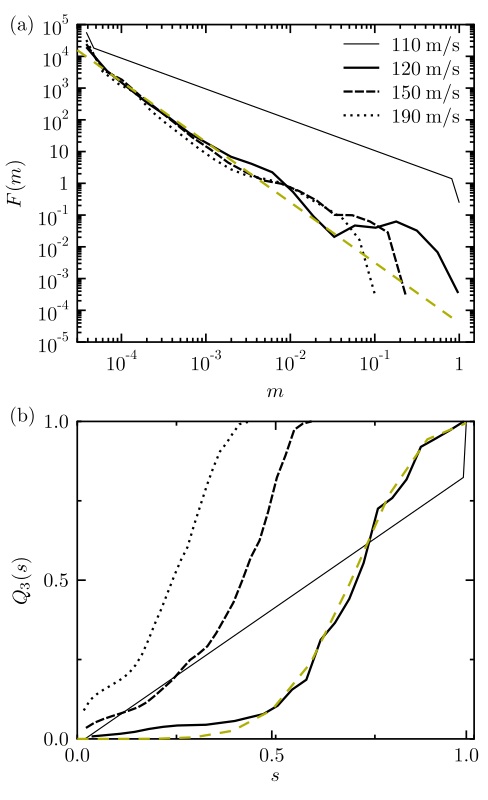}
   \caption{\label{fig:mdist}(a) Fragment mass distribution for
     different initial velocities. The straight line corresponds to a
     power-law with exponent -1.9 (b) Fragment size distribution
     weighted by mass for initial velocities with identical legend as above.}
 \end{figure}

 One of the first and still most important characterizations for
 fragmentation processes are fragment mass distributions.
 Experimental and numerical studies on fragmentation show that the
 mass distribution follow a power law in the range of small fragments,
 whose exponent depends on the fragmentation mechanisms, while the
 mass distribution for large fragments is usually represented by an
 exponential cut-off of the power law.  The fragment mass
 distributions that are obtained from our three-dimensional
 simulations are given in Fig.~\ref{fig:mdist}(a) for different impact
 velocities $v_i$.  Here $F(m)$ represents the probability density of
 finding a fragment with mass $m$ between $m$ and $m + \Delta m$.
 Where $m$ is the fragment mass normalized by the total mass of the
 sphere $M_{tot}$. The values are averaged over 36 simulations,
 changing the random breaking thresholds and randomly rotating the
 sample to obtain different impact points. For velocities below the
 critical velocity $v_{th}$ of our model, the fragment mass
 distribution shows a peak at low fragment masses, corresponding to
 some small fragments. The pronounced isolated peaks near the total
 mass of the system correspond to large damaged, but still unbroken
 system (see also Figs.~\ref{fig:final_cracks}(a) and (b)).  Fragments
 at intermediate mass range are not found for small initial
 velocities. At and above $v_{th}$, the fragment mass distribution
 exhibits a power law dependence for intermediate masses, $F(m) \sim
 m^{-\tau}$, (dashed line in Fig.~\ref{fig:mdist}(a)) with $\tau = 1.9
 \pm 0.2$ \cite{TURCOTTE1986, Linna2005}, and a broad maximum can be
 observed in the histogram for large fragments, indicating that these
 fragments have their origin in mechanisms distinct from the ones that
 form small fragments.  Fig.~\ref{fig:mdist}(b) shows the cumulative
 size distribution of the fragments weighted by mass, $Q_3$, for the
 same velocities.  $Q_3$ is calculated by summing the mass of all the
 fragments smaller then a given size $s$.  The size of a fragment is
 estimated as the diameter of a sphere with identical mass, the values
 are normalized by the sample diameter. By this representation the
 large fragments are better resolved. We can see that the shape of the
 size distribution for large fragments can be described by a
 two-parameter Weibull distribution, $Q_3(s) = 1 - \exp
 [-\left(s/s_c\right)^{k_s}]$ (dashed line in Fig.~\ref{fig:mdist}(b),
 with $s_c=0.75$ and $k_s=5.8$).  The Weibull distribution is used
 here since it has been empirically found to describe a large number
 of fracture experiments, specially for brittle materials
 \cite{Lu2002}.  With increasing initial velocity, the average
 fragment size shifts towards smaller values, also in agreement with
 experimental findings from Refs.~\cite{Antonyuk2006,Cheong2004}.

\begin{figure}[ht]
   \includegraphics[width=8.5cm, bb = 0 0 488 488]{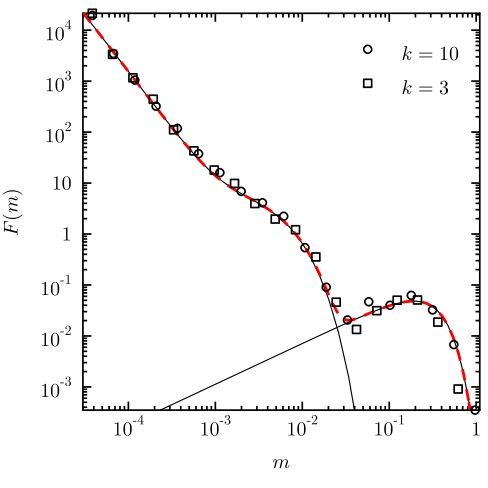}
   \caption{\label{fig:fm_disorder}Fragment mass distribution at
     $v=122.5\text{ m/s}$ for different disorder in the bond breaking
     thresholds. The solid lines correspond to a power law with an
     exponential cuttoff for lower masses and the Weibull distribution for large
     masses (Eq.~(\ref{eq:btd}))}
 \end{figure}

 The local maximum in the fragment mass distribution for large
 fragments represents those fragments, that are formed by the
 meridional cracks. As we can observe from Fig.~\ref{fig:fm_disorder},
 the fragment mass distribution is independent of the amount of
 disorder or material that the specimen is composed of ($k$ is in the
 breaking thresholds distributions in Eq.~(\ref{eq:btd})). Near the
 critical velocity $v_{th}$ we can identify two main parts in the
 fragment mass distribution. For $m$ up to around $1/40$
 (approximately 550 elements), the power law $F(m) \sim m^{-\tau}
 f(m/\bar{m}_o)$ with the cutoff function $f(m/\bar{m}_o)$ containing
 an exponential component $\exp\left(-m/\bar{m}_o\right)$ can be used like in
 Ref. \cite{Wittel2005}. However, for larger $m$, $F(m)$ can also be
 described by a two-parameter Weibull distribution
 \begin{equation}\label{eq:btd}
   F(m) \sim \left(\frac{k_l}{\bar{m}_l}\right)
   \left( \frac{m}{\bar{m}_l}\right)^{k_l-1}
   \exp\left[ -\left( \frac{m}{\bar{m}_l} \right)^{k_l}\right].
 \end{equation}
 The dashed line in Fig.~\ref{fig:fm_disorder} corresponds to a fit to
 the data using $\bar{m}_o=0.004 \pm 0.001$, $\bar{m}_l=0.3 \pm 0.02$
 and $k_l=1.9 \pm 0.1$.  The good quality of the fit allows for a
 better estimation of the exponent of the power-law distribution in
 the small fragment mass range $\tau=2.2 \pm 0.02$.

For the material parameters used in our calculations the primary
cracks have an angular distribution with an average separation from 45
to 60 degrees. Therefore the mass of a fragment resulting from
these plane meridional cracks is of the order of 10\% of the sample
mass, although typically only two to four cracks actually reach the
surface breaking the material. This estimate corresponds to the range
of masses that present the broad peak in the fragment mass
distribution. This feature of the mass distribution function is not
observed in the results of 2D simulations \cite{Kun1999, Behera2005}
or 3D simulations of shell fragmentation \cite{Wittel2004, Wittel2005},
where obviously meridional cracks can not exist.

 \section{Conclusions}
 We studied a brittle, disordered fragmenting solid sphere. We
 performed 3D DEM simulations with 3D beam-truss elements for the
 particle cohesion. Due to this computationally more laborious
 approach as compared to previous works, we were able to obtain a
 clearer picture of the fragmentation process, the evolution of
 fragmentation mechanisms, and its consequences for the fragment mass
 distribution. To get a clearer insight into the fracture initiation,
 we used continuum solutions for the stress field, obtained by the
 Finite Element Method. We were able to show, that 2D simulations for
 fragmenting systems are not capable of capturing fragmentation by
 meridional cracks, that are the primary cracking mechanism. We found
 that cracks form inside the sample in the region above a compressive
 cone long time before they are experimentally observable from the
 outside, if at all. They grow to form meridional fracture planes that
 result in a small number of large wedge shaped fragments, typically
 two to four. The increasing tensile radial and circumferential stress
 in the ring-shaped region above the contact plane gives rise to
 meridional cracks. The decrease in the angular separation between
 these cracks could be explained by the Mott fragmentation model. Some
 of these cracks grow to form the meridional fracture planes that
 break the material in a small number of large fragments, and it is
 only then, that they become visible in experiments.

 The resulting mass distribution of the fragments presents a power law
 regime for small fragments and a broad peak for large fragments that
 can be fitted with a two-parameter Weibull distribution, in agreement
 with experimental results \cite{Salman2002, Lu2002, Cheong2004,
   Antonyuk2006}. The fragment mass distribution is quite robust,
 independent on the macroscopic material properties such as material
 strength and disorder distribution. Only the large fragment range of
 the mass distribution happens to be energy dependent, due to
 additional fragmentation processes that arise as one increases the
 impact energy.

Even though our results are valid for various materials with disorder,
they are limited to the class of brittle, heterogeneous media. Extensions
to ductile materials are in progress. Another class of interesting
questions deal with the problem of size effects, the influence of
polydisperse particles or the stiffness contrast of particles and
beam-elements. The ability of the model to reproduce well defined
crack planes also opens up the possibility to study other crack propagation
problems in 3D. For technological applications questions about the
influence of target shapes and the optimization potential to obtain
desired fragment size distributions or to reduce impact energies are of
broad interest.

\section{Acknowledgments}

We thank the German Federation of Industrial Research Associations
"Otto von Guericke" e.V. (AiF) for financial support, under grant
14516N from the German Federal Ministry of Economics and Technology
(BMWI). H.\ J.\ Herrmann thanks the Max Planck Prize. F.\ Kun was
supported by OTKA T049209. We thank Dr. Jan Bl\"omer and Prof. Jos\'e
Andrade Soares Jr. for helpfull discussions.


 \bibliography{impact}

\begin{thebibliography}{52}
\expandafter\ifx\csname natexlab\endcsname\relax\def\natexlab#1{#1}\fi
\expandafter\ifx\csname bibnamefont\endcsname\relax
  \def\bibnamefont#1{#1}\fi
\expandafter\ifx\csname bibfnamefont\endcsname\relax
  \def\bibfnamefont#1{#1}\fi
\expandafter\ifx\csname citenamefont\endcsname\relax
  \def\citenamefont#1{#1}\fi
\expandafter\ifx\csname url\endcsname\relax
  \def\url#1{\texttt{#1}}\fi
\expandafter\ifx\csname urlprefix\endcsname\relax\def\urlprefix{URL }\fi
\providecommand{\bibinfo}[2]{#2}
\providecommand{\eprint}[2][]{\url{#2}}

\bibitem[{\citenamefont{Herrmann and Roux}(1990)}]{hans:book}
\bibinfo{editor}{\bibfnamefont{H.~J.} \bibnamefont{Herrmann}} \bibnamefont{and}
  \bibinfo{editor}{\bibfnamefont{S.}~\bibnamefont{Roux}}, eds.,
  \emph{\bibinfo{title}{Statistical Models for the Fracture of Disordered
  Media}} (\bibinfo{publisher}{North-Holland}, \bibinfo{address}{Amsterdam},
  \bibinfo{year}{1990}).

\bibitem[{\citenamefont{{\AA}str{\"o}m}(2006)}]{Astrom2006}
\bibinfo{author}{\bibfnamefont{J.~A.} \bibnamefont{{\AA}str{\"o}m}},
  \bibinfo{journal}{Adv. Phys.} \textbf{\bibinfo{volume}{55}},
  \bibinfo{pages}{247} (\bibinfo{year}{2006}).

\bibitem[{\citenamefont{Gilvarry and Bergstrom}(1961)}]{GILVARRY1961a}
\bibinfo{author}{\bibfnamefont{J.~J.} \bibnamefont{Gilvarry}} \bibnamefont{and}
  \bibinfo{author}{\bibfnamefont{B.~H.} \bibnamefont{Bergstrom}},
  \bibinfo{journal}{J. Appl. Phys.} \textbf{\bibinfo{volume}{32}},
  \bibinfo{pages}{400} (\bibinfo{year}{1961}).

\bibitem[{\citenamefont{Gilvarry and Bergstrom}(1962)}]{GILVARRY1962}
\bibinfo{author}{\bibfnamefont{J.~J.} \bibnamefont{Gilvarry}} \bibnamefont{and}
  \bibinfo{author}{\bibfnamefont{B.~H.} \bibnamefont{Bergstrom}},
  \bibinfo{journal}{J. Appl. Phys.} \textbf{\bibinfo{volume}{33}},
  \bibinfo{pages}{3211} (\bibinfo{year}{1962}).

\bibitem[{\citenamefont{Arbiter et~al.}(1969)\citenamefont{Arbiter, Harris, and
  Stamboltzis}}]{ARBITER1969}
\bibinfo{author}{\bibfnamefont{N.}~\bibnamefont{Arbiter}},
  \bibinfo{author}{\bibfnamefont{C.~C.} \bibnamefont{Harris}},
  \bibnamefont{and} \bibinfo{author}{\bibfnamefont{G.~A.}
  \bibnamefont{Stamboltzis}}, \bibinfo{journal}{T. Soc. Min. Eng.}
  \textbf{\bibinfo{volume}{244}}, \bibinfo{pages}{118} (\bibinfo{year}{1969}).

\bibitem[{\citenamefont{Andrews and Kim}(1998)}]{Andrews1998}
\bibinfo{author}{\bibfnamefont{E.~W.} \bibnamefont{Andrews}} \bibnamefont{and}
  \bibinfo{author}{\bibfnamefont{K.~S.} \bibnamefont{Kim}},
  \bibinfo{journal}{Mech. Mater.} \textbf{\bibinfo{volume}{29}},
  \bibinfo{pages}{161} (\bibinfo{year}{1998}).

\bibitem[{\citenamefont{Tomas et~al.}(1999)\citenamefont{Tomas, Schreier,
  Groger, and Ehlers}}]{Tomas1999}
\bibinfo{author}{\bibfnamefont{J.}~\bibnamefont{Tomas}},
  \bibinfo{author}{\bibfnamefont{M.}~\bibnamefont{Schreier}},
  \bibinfo{author}{\bibfnamefont{T.}~\bibnamefont{Groger}}, \bibnamefont{and}
  \bibinfo{author}{\bibfnamefont{S.}~\bibnamefont{Ehlers}},
  \bibinfo{journal}{Powder Technol.} \textbf{\bibinfo{volume}{105}},
  \bibinfo{pages}{39} (\bibinfo{year}{1999}).

\bibitem[{\citenamefont{Majzoub and Chaudhri}(2000)}]{Majzoub2000}
\bibinfo{author}{\bibfnamefont{R.}~\bibnamefont{Majzoub}} \bibnamefont{and}
  \bibinfo{author}{\bibfnamefont{M.~M.} \bibnamefont{Chaudhri}},
  \bibinfo{journal}{Philos. Mag. Lett.} \textbf{\bibinfo{volume}{80}},
  \bibinfo{pages}{387} (\bibinfo{year}{2000}).

\bibitem[{\citenamefont{Chau et~al.}(2000)\citenamefont{Chau, Wei, Wong, and
  Yu}}]{Chau2000}
\bibinfo{author}{\bibfnamefont{K.~T.} \bibnamefont{Chau}},
  \bibinfo{author}{\bibfnamefont{X.~X.} \bibnamefont{Wei}},
  \bibinfo{author}{\bibfnamefont{R.~H.~C.} \bibnamefont{Wong}},
  \bibnamefont{and} \bibinfo{author}{\bibfnamefont{T.~X.} \bibnamefont{Yu}},
  \bibinfo{journal}{Mech. Mater.} \textbf{\bibinfo{volume}{32}},
  \bibinfo{pages}{543} (\bibinfo{year}{2000}).

\bibitem[{\citenamefont{Salman et~al.}(2002)\citenamefont{Salman, Biggs, Fu,
  Angyal, Szabo, and Hounslow}}]{Salman2002}
\bibinfo{author}{\bibfnamefont{A.~D.} \bibnamefont{Salman}},
  \bibinfo{author}{\bibfnamefont{C.~A.} \bibnamefont{Biggs}},
  \bibinfo{author}{\bibfnamefont{J.}~\bibnamefont{Fu}},
  \bibinfo{author}{\bibfnamefont{I.}~\bibnamefont{Angyal}},
  \bibinfo{author}{\bibfnamefont{M.}~\bibnamefont{Szabo}}, \bibnamefont{and}
  \bibinfo{author}{\bibfnamefont{M.~J.} \bibnamefont{Hounslow}},
  \bibinfo{journal}{Powder Technol.} \textbf{\bibinfo{volume}{128}},
  \bibinfo{pages}{36} (\bibinfo{year}{2002}).

\bibitem[{\citenamefont{Wu et~al.}(2004)\citenamefont{Wu, Chau, and
  Yu}}]{Wu2004}
\bibinfo{author}{\bibfnamefont{S.~Z.} \bibnamefont{Wu}},
  \bibinfo{author}{\bibfnamefont{K.~T.} \bibnamefont{Chau}}, \bibnamefont{and}
  \bibinfo{author}{\bibfnamefont{T.~X.} \bibnamefont{Yu}},
  \bibinfo{journal}{Powder Technol.} \textbf{\bibinfo{volume}{143-4}},
  \bibinfo{pages}{41} (\bibinfo{year}{2004}).

\bibitem[{\citenamefont{Schubert et~al.}(2005)\citenamefont{Schubert, Khanal,
  and Tomas}}]{Schubert2005}
\bibinfo{author}{\bibfnamefont{W.}~\bibnamefont{Schubert}},
  \bibinfo{author}{\bibfnamefont{M.}~\bibnamefont{Khanal}}, \bibnamefont{and}
  \bibinfo{author}{\bibfnamefont{J.}~\bibnamefont{Tomas}},
  \bibinfo{journal}{Int. J. Miner. Process.} \textbf{\bibinfo{volume}{75}},
  \bibinfo{pages}{41} (\bibinfo{year}{2005}).

\bibitem[{\citenamefont{Antonyuk et~al.}(2006)\citenamefont{Antonyuk, Khanal,
  Tomas, Heinrich, and Morl}}]{Antonyuk2006}
\bibinfo{author}{\bibfnamefont{S.}~\bibnamefont{Antonyuk}},
  \bibinfo{author}{\bibfnamefont{M.}~\bibnamefont{Khanal}},
  \bibinfo{author}{\bibfnamefont{J.}~\bibnamefont{Tomas}},
  \bibinfo{author}{\bibfnamefont{S.}~\bibnamefont{Heinrich}}, \bibnamefont{and}
  \bibinfo{author}{\bibfnamefont{L.}~\bibnamefont{Morl}},
  \bibinfo{journal}{Chem. Eng. Process.} \textbf{\bibinfo{volume}{45}},
  \bibinfo{pages}{838} (\bibinfo{year}{2006}).

\bibitem[{\citenamefont{Potapov et~al.}(1995)\citenamefont{Potapov, Hopkins,
  and Campbell}}]{POTAPOV1995}
\bibinfo{author}{\bibfnamefont{A.~V.} \bibnamefont{Potapov}},
  \bibinfo{author}{\bibfnamefont{M.~A.} \bibnamefont{Hopkins}},
  \bibnamefont{and} \bibinfo{author}{\bibfnamefont{C.~S.}
  \bibnamefont{Campbell}}, \bibinfo{journal}{Int. J. Mod. Phys. C.}
  \textbf{\bibinfo{volume}{6}}, \bibinfo{pages}{371} (\bibinfo{year}{1995}).

\bibitem[{\citenamefont{Potapov and Campbell}(1996)}]{Potapov1996}
\bibinfo{author}{\bibfnamefont{A.~V.} \bibnamefont{Potapov}} \bibnamefont{and}
  \bibinfo{author}{\bibfnamefont{C.~S.} \bibnamefont{Campbell}},
  \bibinfo{journal}{Int. J. Mod. Phys. C.} \textbf{\bibinfo{volume}{7}},
  \bibinfo{pages}{717} (\bibinfo{year}{1996}).

\bibitem[{\citenamefont{Potapov and Campbell}(1997)}]{Potapov1997}
\bibinfo{author}{\bibfnamefont{A.~V.} \bibnamefont{Potapov}} \bibnamefont{and}
  \bibinfo{author}{\bibfnamefont{C.~S.} \bibnamefont{Campbell}},
  \bibinfo{journal}{Powder Technol.} \textbf{\bibinfo{volume}{93}},
  \bibinfo{pages}{13} (\bibinfo{year}{1997}).

\bibitem[{\citenamefont{Thornton et~al.}(1996)\citenamefont{Thornton, Yin, and
  Adams}}]{Thornton1996}
\bibinfo{author}{\bibfnamefont{C.}~\bibnamefont{Thornton}},
  \bibinfo{author}{\bibfnamefont{K.~K.} \bibnamefont{Yin}}, \bibnamefont{and}
  \bibinfo{author}{\bibfnamefont{M.~J.} \bibnamefont{Adams}},
  \bibinfo{journal}{J. Phys. D. Appl. Phys.} \textbf{\bibinfo{volume}{29}},
  \bibinfo{pages}{424} (\bibinfo{year}{1996}).

\bibitem[{\citenamefont{Kun and Herrmann}(1999)}]{Kun1999}
\bibinfo{author}{\bibfnamefont{F.}~\bibnamefont{Kun}} \bibnamefont{and}
  \bibinfo{author}{\bibfnamefont{H.~J.} \bibnamefont{Herrmann}},
  \bibinfo{journal}{Phys. Rev. E.} \textbf{\bibinfo{volume}{59}},
  \bibinfo{pages}{2623} (\bibinfo{year}{1999}).

\bibitem[{\citenamefont{Thornton et~al.}(1999)\citenamefont{Thornton, Ciomocos,
  and Adams}}]{Thornton1999}
\bibinfo{author}{\bibfnamefont{C.}~\bibnamefont{Thornton}},
  \bibinfo{author}{\bibfnamefont{M.~T.} \bibnamefont{Ciomocos}},
  \bibnamefont{and} \bibinfo{author}{\bibfnamefont{M.~J.} \bibnamefont{Adams}},
  \bibinfo{journal}{Powder Technol.} \textbf{\bibinfo{volume}{105}},
  \bibinfo{pages}{74} (\bibinfo{year}{1999}).

\bibitem[{\citenamefont{Khanal et~al.}(2004)\citenamefont{Khanal, Schubert, and
  Tomas}}]{Khanal2004}
\bibinfo{author}{\bibfnamefont{M.}~\bibnamefont{Khanal}},
  \bibinfo{author}{\bibfnamefont{W.}~\bibnamefont{Schubert}}, \bibnamefont{and}
  \bibinfo{author}{\bibfnamefont{J.}~\bibnamefont{Tomas}},
  \bibinfo{journal}{Granul. Matter.} \textbf{\bibinfo{volume}{5}},
  \bibinfo{pages}{177} (\bibinfo{year}{2004}).

\bibitem[{\citenamefont{Behera et~al.}(2005)\citenamefont{Behera, Kun,
  McNamara, and Herrmann}}]{Behera2005}
\bibinfo{author}{\bibfnamefont{B.}~\bibnamefont{Behera}},
  \bibinfo{author}{\bibfnamefont{F.}~\bibnamefont{Kun}},
  \bibinfo{author}{\bibfnamefont{S.}~\bibnamefont{McNamara}}, \bibnamefont{and}
  \bibinfo{author}{\bibfnamefont{H.~J.} \bibnamefont{Herrmann}},
  \bibinfo{journal}{J. Phys-condens. Mat.} \textbf{\bibinfo{volume}{17}},
  \bibinfo{pages}{S2439} (\bibinfo{year}{2005}).

\bibitem[{\citenamefont{Herrmann et~al.}(2006)\citenamefont{Herrmann, Wittel,
  and Kun}}]{Herrmann2006}
\bibinfo{author}{\bibfnamefont{H.~J.} \bibnamefont{Herrmann}},
  \bibinfo{author}{\bibfnamefont{F.~K.} \bibnamefont{Wittel}},
  \bibnamefont{and} \bibinfo{author}{\bibfnamefont{F.}~\bibnamefont{Kun}},
  \bibinfo{journal}{Physica A} \textbf{\bibinfo{volume}{371}},
  \bibinfo{pages}{59} (\bibinfo{year}{2006}).

\bibitem[{\citenamefont{Kun and Herrmann}(1996{\natexlab{a}})}]{Kun1996}
\bibinfo{author}{\bibfnamefont{F.}~\bibnamefont{Kun}} \bibnamefont{and}
  \bibinfo{author}{\bibfnamefont{H.~J.} \bibnamefont{Herrmann}},
  \bibinfo{journal}{Int. J. Mod. Phys. C.} \textbf{\bibinfo{volume}{7}},
  \bibinfo{pages}{837} (\bibinfo{year}{1996}{\natexlab{a}}).

\bibitem[{\citenamefont{Kun and Herrmann}(1996{\natexlab{b}})}]{Kun1996a}
\bibinfo{author}{\bibfnamefont{F.}~\bibnamefont{Kun}} \bibnamefont{and}
  \bibinfo{author}{\bibfnamefont{H.~J.} \bibnamefont{Herrmann}},
  \bibinfo{journal}{Comput. Method. Appl. M.} \textbf{\bibinfo{volume}{138}},
  \bibinfo{pages}{3} (\bibinfo{year}{1996}{\natexlab{b}}).

\bibitem[{\citenamefont{Diehl et~al.}(2000)\citenamefont{Diehl, Carmona,
  Araripe, Andrade, and Farias}}]{Diehl2000}
\bibinfo{author}{\bibfnamefont{A.}~\bibnamefont{Diehl}},
  \bibinfo{author}{\bibfnamefont{H.~A.} \bibnamefont{Carmona}},
  \bibinfo{author}{\bibfnamefont{L.~E.} \bibnamefont{Araripe}},
  \bibinfo{author}{\bibfnamefont{J.~S.} \bibnamefont{Andrade}},
  \bibnamefont{and} \bibinfo{author}{\bibfnamefont{G.~A.}
  \bibnamefont{Farias}}, \bibinfo{journal}{Phys. Rev. E.}
  \textbf{\bibinfo{volume}{62}}, \bibinfo{pages}{4742} (\bibinfo{year}{2000}).

\bibitem[{\citenamefont{{\AA}str{\"o}m
  et~al.}(2000)\citenamefont{{\AA}str{\"o}m, Holian, and Timonen}}]{Astrom2000}
\bibinfo{author}{\bibfnamefont{J.~A.} \bibnamefont{{\AA}str{\"o}m}},
  \bibinfo{author}{\bibfnamefont{B.~L.} \bibnamefont{Holian}},
  \bibnamefont{and} \bibinfo{author}{\bibfnamefont{J.}~\bibnamefont{Timonen}},
  \bibinfo{journal}{Phys. Rev. Lett.} \textbf{\bibinfo{volume}{84}},
  \bibinfo{pages}{3061} (\bibinfo{year}{2000}).

\bibitem[{\citenamefont{Oddershede et~al.}(1993)\citenamefont{Oddershede,
  Dimon, and Bohr}}]{ODDERSHEDE1993}
\bibinfo{author}{\bibfnamefont{L.}~\bibnamefont{Oddershede}},
  \bibinfo{author}{\bibfnamefont{P.}~\bibnamefont{Dimon}}, \bibnamefont{and}
  \bibinfo{author}{\bibfnamefont{J.}~\bibnamefont{Bohr}},
  \bibinfo{journal}{Phys. Rev. Lett.} \textbf{\bibinfo{volume}{71}},
  \bibinfo{pages}{3107} (\bibinfo{year}{1993}).

\bibitem[{\citenamefont{Meibom and Balslev}(1996)}]{Meibom1996}
\bibinfo{author}{\bibfnamefont{A.}~\bibnamefont{Meibom}} \bibnamefont{and}
  \bibinfo{author}{\bibfnamefont{I.}~\bibnamefont{Balslev}},
  \bibinfo{journal}{Phys. Rev. Lett.} \textbf{\bibinfo{volume}{76}},
  \bibinfo{pages}{2492} (\bibinfo{year}{1996}).

\bibitem[{\citenamefont{Lu et~al.}(2002)\citenamefont{Lu, Danzer, and
  Fischer}}]{Lu2002}
\bibinfo{author}{\bibfnamefont{C.~S.} \bibnamefont{Lu}},
  \bibinfo{author}{\bibfnamefont{R.}~\bibnamefont{Danzer}}, \bibnamefont{and}
  \bibinfo{author}{\bibfnamefont{F.~D.} \bibnamefont{Fischer}},
  \bibinfo{journal}{Phys. Rev. E.} \textbf{\bibinfo{volume}{65}},
  \bibinfo{pages}{067102} (\bibinfo{year}{2002}).

\bibitem[{\citenamefont{Cheong et~al.}(2004)\citenamefont{Cheong, Reynolds,
  Salman, and Hounslow}}]{Cheong2004}
\bibinfo{author}{\bibfnamefont{Y.~S.} \bibnamefont{Cheong}},
  \bibinfo{author}{\bibfnamefont{G.~K.} \bibnamefont{Reynolds}},
  \bibinfo{author}{\bibfnamefont{A.~D.} \bibnamefont{Salman}},
  \bibnamefont{and} \bibinfo{author}{\bibfnamefont{M.~J.}
  \bibnamefont{Hounslow}}, \bibinfo{journal}{Int. J. Miner. Process.}
  \textbf{\bibinfo{volume}{74}}, \bibinfo{pages}{S227} (\bibinfo{year}{2004}).

\bibitem[{\citenamefont{Andrews and Kim}(1999)}]{Andrews1999}
\bibinfo{author}{\bibfnamefont{E.~W.} \bibnamefont{Andrews}} \bibnamefont{and}
  \bibinfo{author}{\bibfnamefont{K.~S.} \bibnamefont{Kim}},
  \bibinfo{journal}{Mech. Mater.} \textbf{\bibinfo{volume}{31}},
  \bibinfo{pages}{689} (\bibinfo{year}{1999}).

\bibitem[{\citenamefont{Cundall and Strack}(1979)}]{CUNDALL1979}
\bibinfo{author}{\bibfnamefont{P.~A.} \bibnamefont{Cundall}} \bibnamefont{and}
  \bibinfo{author}{\bibfnamefont{O.~D.~L.} \bibnamefont{Strack}},
  \bibinfo{journal}{Geotechnique.} \textbf{\bibinfo{volume}{29}},
  \bibinfo{pages}{47} (\bibinfo{year}{1979}).

\bibitem[{\citenamefont{Mishra and Thornton}(2001)}]{Mishra2001}
\bibinfo{author}{\bibfnamefont{B.~K.} \bibnamefont{Mishra}} \bibnamefont{and}
  \bibinfo{author}{\bibfnamefont{C.}~\bibnamefont{Thornton}},
  \bibinfo{journal}{Int. J. Miner. Process.} \textbf{\bibinfo{volume}{61}},
  \bibinfo{pages}{225} (\bibinfo{year}{2001}).

\bibitem[{\citenamefont{Thornton and Liu}(2004)}]{Thornton2004}
\bibinfo{author}{\bibfnamefont{C.}~\bibnamefont{Thornton}} \bibnamefont{and}
  \bibinfo{author}{\bibfnamefont{L.~F.} \bibnamefont{Liu}},
  \bibinfo{journal}{Powder Technol.} \textbf{\bibinfo{volume}{143-4}},
  \bibinfo{pages}{110} (\bibinfo{year}{2004}).

\bibitem[{\citenamefont{Thornton et~al.}(2004)\citenamefont{Thornton, Ciomocos,
  and Adams}}]{Thornton2004a}
\bibinfo{author}{\bibfnamefont{C.}~\bibnamefont{Thornton}},
  \bibinfo{author}{\bibfnamefont{M.~T.} \bibnamefont{Ciomocos}},
  \bibnamefont{and} \bibinfo{author}{\bibfnamefont{M.~J.} \bibnamefont{Adams}},
  \bibinfo{journal}{Powder Technol.} \textbf{\bibinfo{volume}{140}},
  \bibinfo{pages}{258} (\bibinfo{year}{2004}).

\bibitem[{\citenamefont{Potyondy and Cundall}(2004)}]{Potyondy2004}
\bibinfo{author}{\bibfnamefont{D.~O.} \bibnamefont{Potyondy}} \bibnamefont{and}
  \bibinfo{author}{\bibfnamefont{P.~A.} \bibnamefont{Cundall}},
  \bibinfo{journal}{Int. J. Rock. Mech. Min.} \textbf{\bibinfo{volume}{41}},
  \bibinfo{pages}{1329} (\bibinfo{year}{2004}).

\bibitem[{\citenamefont{D'Addetta and Ramm}(2006)}]{DAddetta2006}
\bibinfo{author}{\bibfnamefont{G.~A.} \bibnamefont{D'Addetta}}
  \bibnamefont{and} \bibinfo{author}{\bibfnamefont{E.}~\bibnamefont{Ramm}},
  \bibinfo{journal}{Granul. Matter.} \textbf{\bibinfo{volume}{8}},
  \bibinfo{pages}{159} (\bibinfo{year}{2006}).

\bibitem[{\citenamefont{Carmona et~al.}(2007)\citenamefont{Carmona, Kun,
  {Andrade Jr}, and Herrmann}}]{Carmona2007}
\bibinfo{author}{\bibfnamefont{H.~A.} \bibnamefont{Carmona}},
  \bibinfo{author}{\bibfnamefont{F.}~\bibnamefont{Kun}},
  \bibinfo{author}{\bibfnamefont{J.~S.} \bibnamefont{{Andrade Jr}}},
  \bibnamefont{and} \bibinfo{author}{\bibfnamefont{H.~J.}
  \bibnamefont{Herrmann}}, \bibinfo{journal}{Phys. Rev. E.}
  \textbf{\bibinfo{volume}{75}}, \bibinfo{pages}{046115}
  (\bibinfo{year}{2007}).

\bibitem[{\citenamefont{Landau and Lifshitz}(1986)}]{Landau1986}
\bibinfo{author}{\bibfnamefont{L.~D.} \bibnamefont{Landau}} \bibnamefont{and}
  \bibinfo{author}{\bibfnamefont{E.~M.} \bibnamefont{Lifshitz}},
  \emph{\bibinfo{title}{Theory of Elasticity}}, vol.~\bibinfo{volume}{7} of
  \emph{\bibinfo{series}{Course of Theoretical Physics}}
  (\bibinfo{publisher}{Butterworth-Heinemann, London}, \bibinfo{year}{1986}),
  \bibinfo{edition}{3rd} ed.

\bibitem[{\citenamefont{Herrmann et~al.}(1989)\citenamefont{Herrmann, Hansen,
  and Roux}}]{HERRMANN1989}
\bibinfo{author}{\bibfnamefont{H.~J.} \bibnamefont{Herrmann}},
  \bibinfo{author}{\bibfnamefont{A.}~\bibnamefont{Hansen}}, \bibnamefont{and}
  \bibinfo{author}{\bibfnamefont{S.}~\bibnamefont{Roux}},
  \bibinfo{journal}{Phys. Rev. B.} \textbf{\bibinfo{volume}{39}},
  \bibinfo{pages}{637} (\bibinfo{year}{1989}).

\bibitem[{\citenamefont{P{\"o}schel and Schwager}(2005)}]{Poschel2005}
\bibinfo{author}{\bibfnamefont{T.}~\bibnamefont{P{\"o}schel}} \bibnamefont{and}
  \bibinfo{author}{\bibfnamefont{T.}~\bibnamefont{Schwager}},
  \emph{\bibinfo{title}{Computational Granular Dynamics: Models and
  Algorithms}} (\bibinfo{publisher}{Springer-Verlag Berlin Heidelberg New
  York}, \bibinfo{year}{2005}).

\bibitem[{\citenamefont{D'Addetta et~al.}(2001)\citenamefont{D'Addetta, Kun,
  Ramm, and Herrmann}}]{DAddetta2001}
\bibinfo{author}{\bibfnamefont{G.~A.} \bibnamefont{D'Addetta}},
  \bibinfo{author}{\bibfnamefont{F.}~\bibnamefont{Kun}},
  \bibinfo{author}{\bibfnamefont{E.}~\bibnamefont{Ramm}}, \bibnamefont{and}
  \bibinfo{author}{\bibfnamefont{H.~J.} \bibnamefont{Herrmann}}, in
  \emph{\bibinfo{booktitle}{Continuous and Discontinuous Modelling of
  Cohesive-Frictional Materials}}, edited by
  \bibinfo{editor}{\bibfnamefont{P.}~\bibnamefont{Vermeer}}
  (\bibinfo{publisher}{Springer-Verlag}, \bibinfo{address}{Berlin},
  \bibinfo{year}{2001}), vol. \bibinfo{volume}{568} of
  \emph{\bibinfo{series}{Springer Lecture Notes in Physics}}, pp.
  \bibinfo{pages}{231--258}.

\bibitem[{\citenamefont{Lilliu and Van~Mier}(2003)}]{Lilliu2003}
\bibinfo{author}{\bibfnamefont{G.}~\bibnamefont{Lilliu}} \bibnamefont{and}
  \bibinfo{author}{\bibfnamefont{J.~G.~M.} \bibnamefont{Van~Mier}},
  \bibinfo{journal}{Eng. Fract. Mech.} \textbf{\bibinfo{volume}{70}},
  \bibinfo{pages}{927} (\bibinfo{year}{2003}).

\bibitem[{\citenamefont{Rapaport}(2004)}]{Rapaport2004}
\bibinfo{author}{\bibfnamefont{D.~C.} \bibnamefont{Rapaport}},
  \emph{\bibinfo{title}{The Art of Molecular Dynamics Simulation}}
  (\bibinfo{publisher}{Cambridge University Press, Cambridge},
  \bibinfo{year}{2004}), \bibinfo{edition}{2nd} ed.

\bibitem[{\citenamefont{Bolander and Sukumar}(2005)}]{Bolander2005}
\bibinfo{author}{\bibfnamefont{J.~E.} \bibnamefont{Bolander}} \bibnamefont{and}
  \bibinfo{author}{\bibfnamefont{N.}~\bibnamefont{Sukumar}},
  \bibinfo{journal}{Phys. Rev. B.} \textbf{\bibinfo{volume}{71}},
  \bibinfo{pages}{094106} (\bibinfo{year}{2005}).

\bibitem[{\citenamefont{Yip et~al.}(2006)\citenamefont{Yip, Li, Liao, and
  Bolander}}]{Yip2006}
\bibinfo{author}{\bibfnamefont{M.}~\bibnamefont{Yip}},
  \bibinfo{author}{\bibfnamefont{Z.}~\bibnamefont{Li}},
  \bibinfo{author}{\bibfnamefont{B.~S.} \bibnamefont{Liao}}, \bibnamefont{and}
  \bibinfo{author}{\bibfnamefont{J.~E.} \bibnamefont{Bolander}},
  \bibinfo{journal}{Int. J. Fracture.} \textbf{\bibinfo{volume}{140}},
  \bibinfo{pages}{113} (\bibinfo{year}{2006}).

\bibitem[{\citenamefont{Baram and Herrmann}(2005)}]{Baram2005}
\bibinfo{author}{\bibfnamefont{R.~M.} \bibnamefont{Baram}} \bibnamefont{and}
  \bibinfo{author}{\bibfnamefont{H.~J.} \bibnamefont{Herrmann}},
  \bibinfo{journal}{Phys. Rev. Lett.} \textbf{\bibinfo{volume}{95}},
  \bibinfo{pages}{224303} (\bibinfo{year}{2005}).

\bibitem[{\citenamefont{Mott}(1946)}]{MOTT46}
\bibinfo{author}{\bibfnamefont{N.~F.} \bibnamefont{Mott}},
  \bibinfo{journal}{Proceedings of the Royal Society of London A}
  \textbf{\bibinfo{volume}{189}}, \bibinfo{pages}{300} (\bibinfo{year}{1946}).

\bibitem[{\citenamefont{Wittel et~al.}(2005)\citenamefont{Wittel, Kun,
  Herrmann, and Kroplin}}]{Wittel2005}
\bibinfo{author}{\bibfnamefont{F.~K.} \bibnamefont{Wittel}},
  \bibinfo{author}{\bibfnamefont{F.}~\bibnamefont{Kun}},
  \bibinfo{author}{\bibfnamefont{H.~J.} \bibnamefont{Herrmann}},
  \bibnamefont{and} \bibinfo{author}{\bibfnamefont{B.~H.}
  \bibnamefont{Kroplin}}, \bibinfo{journal}{Phys. Rev. E.}
  \textbf{\bibinfo{volume}{71}}, \bibinfo{pages}{016108}
  (\bibinfo{year}{2005}).

\bibitem[{\citenamefont{Turcotte}(1986)}]{TURCOTTE1986}
\bibinfo{author}{\bibfnamefont{D.~L.} \bibnamefont{Turcotte}},
  \bibinfo{journal}{J. Geophys. Res-solid.} \textbf{\bibinfo{volume}{91}},
  \bibinfo{pages}{1921} (\bibinfo{year}{1986}).

\bibitem[{\citenamefont{Linna et~al.}(2005)\citenamefont{Linna, {\AA}str{\"o}m,
  and Timonen}}]{Linna2005}
\bibinfo{author}{\bibfnamefont{R.~P.} \bibnamefont{Linna}},
  \bibinfo{author}{\bibfnamefont{J.~A.} \bibnamefont{{\AA}str{\"o}m}},
  \bibnamefont{and} \bibinfo{author}{\bibfnamefont{J.}~\bibnamefont{Timonen}},
  \bibinfo{journal}{Phys. Rev. E.} \textbf{\bibinfo{volume}{72}},
  \bibinfo{pages}{015601} (\bibinfo{year}{2005}).

\bibitem[{\citenamefont{Wittel et~al.}(2004)\citenamefont{Wittel, Kun,
  Herrmann, and Kr{\"o}plin}}]{Wittel2004}
\bibinfo{author}{\bibfnamefont{F.}~\bibnamefont{Wittel}},
  \bibinfo{author}{\bibfnamefont{F.}~\bibnamefont{Kun}},
  \bibinfo{author}{\bibfnamefont{H.~J.} \bibnamefont{Herrmann}},
  \bibnamefont{and} \bibinfo{author}{\bibfnamefont{B.~H.}
  \bibnamefont{Kr{\"o}plin}}, \bibinfo{journal}{Phys. Rev. Lett.}
  \textbf{\bibinfo{volume}{93}}, \bibinfo{pages}{035504}
  (\bibinfo{year}{2004}).

\end{thebibliography}
\end{document}